\documentclass[a4paper]{article}
\usepackage{amsmath}
\usepackage[english]{babel}
\usepackage{latexsym}
\usepackage{amssymb}
\usepackage{amscd}

\numberwithin{equation}{section}
\newcommand{\bdm}{\begin{displaymath}}
\newcommand{\edm}{\end{displaymath}}
\newcommand{\bdn}{\begin{eqnarray}}
\newcommand{\edn}{\end{eqnarray}}
\newcommand{\bay}{\begin{array}{c}}
\newcommand{\eay}{\end{array}}
\newcommand{\ben}{\begin{enumerate}}
\newcommand{\een}{\end{enumerate}}
\newcommand{\beq}{\begin{equation}}
\newcommand{\eeq}{\end{equation}}
\newtheorem{lem}{Lemma}[section]
\newtheorem{teo}{Theorem}[section]
\newtheorem{pro}{Proposition}[section]

\newtheorem{cor}{Corollary}[section]

\title{Ionization for Three Dimensional Time-dependent Point Interactions	\\	\mbox{}	\\}
\author{Michele Correggi\footnote{E-mail address: \texttt{correggi@sissa.it}}	\\	\small{International School for Advanced Studies SISSA/ISAS, Trieste, Italy}	\\	\mbox{}	\\	Gianfausto Dell'Antonio\footnote{E-mail address: \texttt{gianfa@sissa.it}}	\\	\small{Centro Linceo Interdisciplinare\footnote{On leave from Dipartimento di Matematica, Universit\`{a} di Roma, ``La Sapienza'', Italy.}, Roma, Italy}	\\	\mbox{}	\\	Rodolfo Figari\footnote{E-mail address: \texttt{figari@na.infn.it}}	\\	\small{Dipartimento di Scienze Fisiche, Universit\`{a} di Napoli ``Federico II''}	\\	\small{ and Sezione INFN, Napoli, Italy}	\\	\mbox{}	\\	Andrea Mantile\footnote{E-mail address: \texttt{andrea.mantile@dma.unina.it}}	\\	\small{Dipartimento di Matematica e Applicazioni}	\\
	\small{Universit\`{a} di Napoli ``Federico II'', Napoli, Italy} 	\\} 
\date{}

\begin{document}

\maketitle

\begin{abstract}
	We study the time evolution of a three dimensional quantum particle under the action of a time-dependent point interaction fixed at the origin. We assume that the ``strength'' of the interaction \( \alpha(t) \) is a periodic function with an arbitrary mean. Under very weak conditions on the Fourier coefficients of \( \alpha(t) \), we prove that there is complete ionization as \( t \rightarrow \infty \), starting from a bound state at time \( t = 0 \). Moreover we prove also that, under the same conditions, all the states of the system are scattering states.
\end{abstract}

\begin{center}
	Ref. SISSA/ISAS preprint 11/2004/FM
\end{center}

\section{Introduction}

We shall study the time evolution of a three dimensional system with time-dependent Hamiltonian given by
	\bdm
		H(t) = H_0 + H_I(t)
	\edm
where the ``perturbation'' \( H_I(t) \) is a zero-range interaction with time-dependent (periodic) ``strength''. In particular we are interested in proving complete ionization of the system as \( t \rightarrow \infty \), starting from an initial condition at \( t = 0 \) given by a bound state of the system. By complete ionization one can mean two different statements. The weaker one is that the survival probability of the bound state, i.e. the square modulus of the scalar product of the state at time \( t  \) with the bound state, goes to zero as \( t \rightarrow \infty \). The stronger one is that every state \( \Psi \) in the Hilbert space of the system is a scattering state (see for example \cite{Enss1,Howl1}) of \( H(t) \), i.e. for every compact set \( S \subset \mathbb{R}^3 \), 
	\bdm
		\lim_{t \rightarrow \infty} \frac{1}{t} \int_0^t d\tau \int_S d^3 \vec{x} \:\: \big| \Psi_{\tau}(\vec{x}) \big|^2 = 0
	\edm
	\( \Psi_t \) denoting the time evolution of the state \( \Psi \). The last statement is related to the absence of eigenvalues of the Floquet operator associated to \( H(t) \) (see \cite{Howl2,Graf1,Yaji1}). 
\newline
The usual way to deal with problems of this kind is by means of time-dependent perturbation theory and Fermi's golden rule, which gives for the survival probability the well known exponential decay for each order \( n \) in the perturbative expansion. On the other hand simple examples of regular perturbations show that the survival probability decays to zero as a power-law (i.e. the limits \( t \rightarrow \infty \) and \( n \rightarrow \infty \) can not be interchanged).  When the perturbation is not small, it is in general very difficult to solve the problem and find the law of decay. Therefore it is interesting to find models in which a non-perturbative solution exists and study the survival probability. In this paper we study one such model, in which \( H_I(t) \) is given by a three dimensional point interaction. We shall see that it is possible to prove asymptotic complete ionization and find a power law decay for the survival probability, under generic condition on the scattering length\footnote{In three dimensions the parameter \( \alpha(t) \) is proportional to the inverse of the scattering length.}.
\newline
The one-dimensional version of the same problem has been widely analyzed in \( \cite{Cost1,Cost2,Cost3,Cost4} \), where complete ionization is proved under a suitable and very weak condition on the Fourier coefficients of the strength of the interaction. We shall see that the same \emph{genericity} condition is also sufficient in three dimensions to have complete ionization of the system.
\newline
From a physical point of view, the model we are going to study is related to the strong laser ionization of Rydberg atoms\footnote{See the discussion contained in \cite{Cost1,Cost5} and references therein.}, showing many features of experimental data. Indeed, despite of the simplicity of the model, as in the one-dimensional case, it is possible to reproduce many effects of multiphoton ionization of excited hydrogen atoms by microwave field, with a good agreement with experiments (see \cite{Cost5}).

\section{The model}

The model we are going to study is a quantum particle subjected to a time-dependent point interaction fixed at the origin in three dimensions, namely a system defined by the time-dependent self-adjoint Hamiltonian \( H_{\alpha(t)} \),
	\bdm
		\mathcal{D}(H_{\alpha(t)}) = \Big\{ \Psi \in L^2(\mathbb{R}^3) \: \big| \: \exists \: q_{\lambda}(t) \in \mathbb{C}, \big(\Psi(\vec{x}) - q_{\lambda}(t) \: \mathcal{G}^{\lambda}(\vec{x}) \big) \in H^2(\mathbb{R}^3), 
	\edm
	\beq
		\big(\Psi - q_{\lambda}(t) \: \mathcal{G}^{\lambda} \big)\big|_{\vec{x} = 0} = \bigg( \alpha(t) + \frac{\sqrt{\lambda}}{4 \pi} \bigg) \: q_{\lambda}(t) \bigg\}
	\eeq
	\beq
	\label{Operator}
		\big( H_{\alpha(t)} + \lambda \big) \Psi = \big( H_0 + \lambda \big) \big( \Psi - q_{\lambda}(t) \: \mathcal{G}^{\lambda} \big)
	\eeq
where \( \lambda \in \mathbb{R} \), \( \lambda > 0 \) and
	\bdm
		\mathcal{G}^{\lambda}(\vec{x} - \vec{x}^{\prime}) = \frac{e^{-\sqrt{\lambda}|\vec{x} - \vec{x}^{\prime}|}}{4\pi |\vec{x}-\vec{x}^{\prime}|}
	\edm
is the Green function of the free Hamiltonian \( H_0 = - \Delta \).
\newline
The operator\footnote{For a general review about point interactions see \cite{Albe1,Bere1} and references therein.} (\ref{Operator}) has absolutely continuous spectrum if \( \alpha(t) \) is positive, while, when \( \alpha(t) < 0 \), there exists exactly one negative eigenvalue \( - (4 \pi \alpha(t))^2 \), with normalized eigenfunction
	\beq
		\varphi_{\alpha(t)}(\vec{x}) \equiv \frac{\sqrt{2 |\alpha(t)|} \: e^{4 \pi \alpha(t) |\vec{x}|}}{|\vec{x}|} 
	\eeq
It is well known (see \cite{Dell1,Dell2,Figar1,Saya1,Yafa1}) that the operator (\ref{Operator}) defines a time propagation \( U(t,s) \) given by a two-parameters unitary family, solving the time-dependent Shr\"{o}dinger equation
	\beq
	\label{Schro}
		i \frac{\partial \Psi_t}{\partial t} = H_{\alpha(t)} \Psi_t
	\eeq
and 
	\beq
	\label{State}
		\Psi_t(\vec{x}) = U(t,s) \: \Psi_s (\vec{x}) = U_0(t-s) \Psi_s (\vec{x}) + i \int_s^t d \tau \: q(\tau) \: U_0(t-\tau ; \vec{x})
	\eeq
where \( U_0(t) = \exp( -iH_0t) \), \( U_0(t;\vec{x}) \) is the kernel associated to the free propagator and the charge \( q(t) \) satisfies a Volterra integral equation for \( t \geq s \),
	\beq
	\label{Equation}
		q(t) + 4 \sqrt{\pi i} \int_s^t d\tau \: \frac{\alpha(\tau) q(\tau)}{\sqrt{t-\tau}} = 4 \sqrt{\pi i} \int_s^t d\tau \: \frac{\big(U_0(\tau) \Psi_s\big)(0)}{\sqrt{t-\tau}}
	\eeq
We are interested in studying complete ionization of system defined by (\ref{Operator}) and (\ref{Schro}), starting from initial conditions 
	\beq
	\label{Initial}
		\Psi_0 (\vec{x}) = \varphi_{\alpha(0)}(\vec{x})
	\eeq
\( \varphi_{\alpha(0)}(\vec{x}) \) being the bound state\footnote{In order to do this analysis we shall require that \( \alpha(0) < 0 \).} of \( H_{\alpha(0)} \). 
\newline
We shall assume that \( \alpha(t) \) is a real periodic continuous function with period \( T \).
\newline
The meaningful parameter of the system is the negative lower bound of \( \alpha(t) \). Indeed, if \( \inf(\alpha(t)) \geq 0 \), the wave operator associated to \( (H_0, H_{\alpha(t)}) \) is unitary (see \cite{Yafa1}) so that any initial state evolves into a scattering state (see also the remark at the end of section 5). Hence we require that
	\beq
	\label{Conditions1}
		\begin{array}{ll}
			1.	&	\alpha(0) < 0 \\
		\eay
	\eeq									
Continuity of \( \alpha(t) \) guarantees that it can be decomposed in a Fourier series, for each \( t \in \mathbb{R}^+ \), and the series converges uniformly on every compact subset of the real line. In terms of the Fourier coefficients of  \( \alpha(t) \), we assume
	\beq
	\label{Conditions2}
		\begin{array}{ll}
			2.	&	\alpha(t) = \displaystyle{\sum_{n \in \mathbb{Z}}} \: \alpha_n \: e^{-i n \omega t} \: , \: \{ \alpha_n \} \in \ell_1(\mathbb{Z}), \:\: \omega = \displaystyle{\frac{2 \pi}{T}}	\\
			\mbox{}	&	\\
			3.	&	\alpha_n = \alpha_{-n}^*	\\
			\mbox{}	&	\\
		\eay
	\eeq	
We start noticing that from (\ref{Equation}) we have
	\beq
		|q(t)| \leq  4 \sqrt{\pi} \: \sup(|\alpha|) \int_0^t d\tau \: \frac{|q(\tau)|}{\sqrt{t-\tau}} + 4 \sqrt{\pi} \int_0^t d\tau \: \frac{ \big| \big( U_0(\tau) \Psi_0 \big) \big|(0)}{\sqrt{t-\tau}}
	\eeq
from which we deduce that \( \eta(t) - |q(t)| \geq 0 \), if \( \eta(t) \) is the unique solution of the equation
	\beq
	\label{Eqmodulus}
		\eta(t) =  4 \sqrt{\pi} \: \sup(|\alpha|) \int_0^t d\tau \: \frac{\eta(\tau)}{\sqrt{t-\tau}} + 4 \sqrt{\pi} \int_0^t d\tau \: \frac{\big| \big(U_0(\tau) \Psi_0 \big) \big|(0)}{\sqrt{t-\tau}}
	\eeq 
Iterating (\ref{Eqmodulus}) once and differentiating we obtain for \( \eta \) the differential equation
	\beq
		\frac{d \eta}{dt} = 16 \pi^{2} \big( \sup(|\alpha|) \big)^2 \eta + 16 \pi^{2} \big| \big(U_0(t) \Psi_0 \big) \big|(0)
	\eeq 
where the inhomogeneous term is finite at each time \( t \) with, at most, an integrable singularity at \( t = 0 \). We conclude that
	\beq
		|q(t)| \leq  \eta(t) \leq C \: e^{16 \pi^{2} (\sup(|\alpha|))^{2} t}
	\eeq 
As a consequence the Laplace transform of \( q(t) \), denoted by
	\bdm
		\tilde{q}(p) \equiv \int_0^{\infty} dt \: e^{-pt} q(t)
	\edm
exists analytic at least for \( \Re(p) > 16 \pi^{2} \big( \sup(|\alpha|) \big)^{2} \). 
\newline
Applying the Laplace transform to equation (\ref{Equation}), one has 
	\beq
	\label{Laplace}
		\tilde{q}(p) = - 4 \pi \sqrt{\frac{i}{p}} \: \sum_{k \in \mathbb{Z}} \: \alpha_k \: \tilde{q}(p+i \omega k) + \tilde{f}(p)
	\eeq
where
	\bdm
		\tilde{f}(p) \equiv \frac{ 2 \sqrt{2 |\alpha(0)|}}{\pi} \: \sqrt{\frac{i}{p}} \: \int_0^{\infty} dt \: e^{-pt} \int_{\mathbb{R}^3} d^3 \vec{k} \: \frac{e^{-ik^2t}}{ k^2 + (4 \pi \alpha(0))^2} =
	\edm
	\bdm
		= 8 \: \sqrt{\frac{2|\alpha(0)|}{ip}} \: \int_0^{\infty} dk \: \frac{k^2}{ (k^2 + (4 \pi \alpha(0))^2)(k^2-ip)} =
	\edm 
	\bdm
		= 4 \pi i \sqrt{\frac{2 |\alpha(0)|}{-ip}} \: \frac{4 \pi \alpha(0) + \sqrt{-ip}}{(4\pi \alpha(0))^2 + ip}
	\edm
and with the choice of the branch cut for the square root along the negative real line: if \( p = \varrho \: e^{i \vartheta} \),
	\beq
	\label{Branch}
		\sqrt{p} = \sqrt{\varrho} \:\: e^{i \vartheta / 2}
	\eeq
with \( -\pi < \vartheta \leq \pi \).
\newline
By unitarity of the evolution (\ref{Schro}), it follows that the Laplace transform of \( q(t) \) is indeed analytic on the open right half plane:

	\begin{pro}
	\label{Analy}
		The Laplace transform of \( q(t) \), solution of (\ref{Equation}), is analytic at least for \( \Re(p) > 0 \). 
	\end{pro}
		
	\emph{Proof:}
		Using the decomposition of the wave function at time \( t \) defined by (\ref{State}), we can write the survival probability in the following way:
		\beq
		\label{Survival}
			\theta(t) \equiv \Big( \varphi_{\alpha(0)} \: , \Psi_t \Big)_{L^2(\mathbb{R}^3)} = \bigg( \varphi_{\alpha(0)} \: , \: e^{-iH_0 t} \varphi_{\alpha(0)} \bigg)_{L^2(\mathbb{R}^3)} +
		\eeq
		\bdm
			+ i \bigg( \varphi_{\alpha(0)}(\vec{x}) \: , \: \int_0^t \: d\tau \: q(\tau) \: U_0(t-\tau ; \vec{x}) \bigg)_{L^2(\mathbb{R}^3)}
		\edm
		Let us define 
		\bdm	
			Z_1(t) \equiv \bigg( \varphi_{\alpha(0)} \: , \: e^{-iH_0 t} \varphi_{\alpha(0)} \bigg)_{L^2(\mathbb{R}^3)}
		\edm
		By the usual dissipative estimate for the free propagator, one has
		\bdm
			\big| Z_1(t) \big| \leq c_1 \: t^{-\frac{3}{2}}
		\edm
		as \( t \rightarrow \infty \) for some constant \( c_1 \in \mathbb{R} \). Hence \( Z_1(t) \) belongs to \( L^1(\mathbb{R}^+) \) and then its Laplace transform \( \tilde{Z}_1(p) \) is analytic at least for \( \Re(p) \geq 0 \).
		\newline
		The second piece of the scalar product is given by
		\bdm
			Z(t) \equiv i \bigg( \varphi_{\alpha(0)}(\vec{x}) \: , \: \int_0^t \: d\tau \: q(\tau) \: U_0(t-\tau ; \vec{x}) \bigg)_{L^2(\mathbb{R}^3)} = 
		\edm
		\bdm
			= i \int_0^t \: d\tau \: q(\tau) \: \bigg( e^{-iH_0(t-\tau)} \varphi_{\alpha(0)} \bigg)(0) 
		\edm
		and taking the Laplace transform of \( Z(t) \), we have
		\bdm
			\tilde{Z}(p) = \tilde{Z}_2(p) \: \tilde{q}(p)
		\edm
		where
		\bdm
			\tilde{Z}_2(p) \equiv - \frac{4 \sqrt{2 \pi |\alpha(0)|}}{4\pi \alpha(0) - \sqrt{-ip}}
		\edm
		is analytic for \( \Re(p) > 0 \) and never equal to \( 0 \), because of condition (\ref{Conditions1}).
		\newline
		Hence the Laplace transform of \( \theta(t) \) is given by
		\bdm
			\tilde{\theta}(p) = \tilde{Z}_1(p) + \tilde{Z}_2(p) \: \tilde{q}(p)
		\edm 
		But \( \theta(t) \) is a bounded function\footnote{Actually \( |\theta(t)| \leq 1 \), since the initial state is normalized.}, because of unitarity of the evolution (\ref{Schro}), and then its Laplace transform is analytic on the open right half plane. The claim then follows from analyticity of \( \tilde{Z}_1(p) \), \( \tilde{Z}_2(p) \) and absence of zeros of \( \tilde{Z}_2(p) \).
		\begin{flushright} 
			\( \Box \)
		\end{flushright}

A well known property of Volterra integral operators, with regular or weakly singular kernel, implies

	\begin{pro}
	\label{Homogeneous}
		The homogeneous equations associated to (\ref{Equation}) has no non-zero solution in \( L^p_{\mathrm{loc}}(\mathbb{R}^+) \), \( 1 \leq p \leq \infty \).
	\end{pro}
		
	\emph{Proof:}
		The proof (see e.g. \cite{Port1}) exploits the fact that the \(n\)-fold iterated kernel is a contraction in any \( L^{p}(0,T_{n}) \) with \( T_{n} \) increasing to infinity for increasing \( n \).
		\begin{flushright} 
			\( \Box \)
		\end{flushright}

In the following sections we shall prove asymptotic complete ionization of the system under generic conditions on \( \alpha(t) \). Although the result does not depend on the sign of the mean \( \alpha_0 \) of \( \alpha(t) \), we have to discuss separately the case \( \alpha_0 < 0 \) and \( \alpha_0 \geq 0 \), because of the slightly different form of equation (\ref{Laplace}).

\section{CASE I: \( \alpha_0 < 0 \)}

Since \( \alpha(0) < 0 \), changing the energy scale, it is always possible to assume that \( \alpha(t) \) satisfies the normalization
	\beq
	\label{Conditions3}
		\begin{array}{ll}
			4.	&	\alpha(0) = \displaystyle{\sum_{n \in \mathbb{Z}}} \: \alpha_n = - \displaystyle{\frac{1}{4 \pi}}	\\ 	
		\eay
	\eeq
Moreover we introduce another condition we shall use later on: let \( \mathcal{T} \) the right shift operator on \( \ell_1(\mathbb{N}) \), i.e.
	\beq
		\big( \mathcal{T} a \big)_n \equiv a_{n+1}
	\eeq
we say that \( \alpha = \{ \alpha_n \} \in \ell_1(\mathbb{Z}) \) is \emph{generic} with respect to \( \mathcal{T} \), if \( \tilde{\alpha} \equiv \{ \alpha_n \}_{n>0} \in \ell_1(\mathbb{N}) \) satisfies the following condition
	\beq
	\label{Genericity}
		e_1 = \big( 1,0,0, \ldots \big) \in \overline{\bigvee_{n=0}^{\infty} \mathcal{T}^n \tilde{\alpha}}
	\eeq
For a detailed discussion of genericity condition see \cite{Cost1}.
\newline
If (\ref{Conditions3}) holds, equation (\ref{Laplace}) becomes (at least for \( \Re(p) > 0 \))
	\beq
	\label{Eq1}
		\tilde{q}(p) =  - \frac{4 \pi}{4 \pi \alpha_0 + \sqrt{-ip}} \: \underset{k \neq 0}{\sum_{k \in \mathbb{Z}}} \: \alpha_k \: \tilde{q}(p+i \omega k) - \frac{2 i \sqrt{2 \pi}}{4 \pi \alpha_0 + \sqrt{-ip}} \frac{1 - \sqrt{-ip}}{1 + ip}
	\eeq
and by Proposition \ref{Analy} its solution is analytic on the open right half plane. In the following section we shall extend the equation (\ref{Eq1}) above to the imaginary axis and study the behavior of the solution there.

\subsection{Behavior on the imaginary axis at \( p \neq 0 \)}

Setting \( q_n(p) \equiv \tilde{q}(p+i \omega n) \), we obtain a sequence of functions on the strip \( \mathcal{I} = \{ p \in  \mathbb{C}, \: 0 \leq \Im(p) < \omega \} \). Setting
	\bdm
		q(p) \equiv \{ q_n(p) \}_{n \in \mathbb{Z}} 
	\edm 
equation (\ref{Eq1}) can be rewritten
	\beq
	\label{Eqr1}
		q(p) = \mathcal{L}(p) \:  q(p) + g(p)
	\eeq
where
	\beq	
	\label{LOperator}
		 \big( \mathcal{L} q \big)_n (p) \equiv - \frac{4 \pi}{4 \pi \alpha_0 + \sqrt{\omega n - ip}} \: \underset{k \neq 0}{\sum_{k \in \mathbb{Z}}} \: \alpha_k \: q_{n+k}(p)
	\eeq
and \( g(p) = \{ g_n(p) \}_{n \in \mathbb{Z}} \) with
	\beq
	\label{Known}
		g_n(p) \equiv - \frac{2 i \sqrt{2 \pi}}{4 \pi \alpha_0 + \sqrt{\omega n -ip}} \frac{1 - \sqrt{\omega n-ip}}{1 + ip - \omega n}
	\eeq
From the explicit expression of the operator (\ref{LOperator}) and (\ref{Known}), it is clear that the coefficients of the equation fails to be analytic on the imaginary axis at \( \bar{p} = ((4 \pi \alpha_0)^2 - \omega \bar{n})i \), for some \( \bar{n} \in \mathbb{Z} \) and then the solution may be singular there.
\newline
Since \( \Im(p) \in [0,\omega) \), one has
	\beq
	\label{Integer}
		\frac{(4 \pi \alpha_0)^2}{\omega} - 1 < \bar{n} \leq \frac{(4 \pi \alpha_0)^2}{\omega}
	\eeq
and then the singularity appears at most in the equation for \( q_{\bar{n}} \) (there is only one integer\footnote{In fact \( \bar{n} \) must be non negative.} which satisfies the previous inequality) at \( \bar{p} = ((4 \pi \alpha_0)^2 - \omega \bar{n})i \).  For instance, if \( \omega > (4 \pi \alpha_0)^2 \), the pole may be at \( \bar{p} = (4 \pi \alpha_0)^2 i \) in the equation for \( q_0 \).
\newline
Actually we have to distinguish the so called (see \cite{Cost1}) resonant case, i.e. when
	\bdm
		(4 \pi \alpha_0)^2 = N \omega
	\edm
for some \( N \in \mathbb{N} \), because in that case we can have a pole only at \( p = 0 \) and then the solution is immediately seen to be analytic on the whole imaginary axis except at most for \( p = 0 \). 
\newline
Let us first consider the behavior of the solution on the imaginary axis for \( p \neq 0, \bar{p} \). We are going to prove that the solution is in fact analytic there. We prove first an important property of the operator \( \mathcal{L} \):

	\begin{pro}
	\label{Compact}
		For \( p \in \mathcal{I} \), \( \Re(p) = 0 \), \( p \neq 0, \bar{p} \), \( \mathcal{L}(p) \) is an analytic operator-valued function and \( \mathcal{L}(p) \) is a compact operator on \( \ell_2(\mathbb{Z}) \).
	\end{pro}

	\emph{Proof:}	
		Analyticity on the imaginary axis for \( p \neq 0, \bar{p} \) easily follows from the explicit expression of the operator. 
		\newline
		Moreover \( \mathcal{L}(p) \) can be written
		\bdm
			\mathcal{L}(p) = b(p) \: \underset{k \neq 0}{\sum_{k \in \mathbb{Z}}} \: \alpha_k \: \mathcal{T}^{n+k}
		\edm
		where \( b(p) \) is the operator
		\bdm
			(b \: q)_n (p) \equiv b_n(p) \: q_n(p) =  - \frac{4 \pi q_n(p)}{4 \pi \alpha_0 + \sqrt{\omega n - ip}}
		\edm
		and \( \mathcal{T} \) is the right shift operator on \( \ell_2(\mathbb{Z}) \).
		\newline
		Since \( \| \mathcal{T} \| = 1 \), the series converges strongly to a bounded operator. Moreover \( b(p) \) is a compact operator on the imaginary axis for \( p \neq 0, \bar{p} \): \( b(p) \) is the norm limit of a sequence of finite rank operators, because \( \lim_{n \rightarrow \infty} b_n(p) = 0 \). Hence the result follows for example from Theorem VI.12 and VI.13 of \cite{Reed1}.
		\begin{flushright} 
			\( \Box \)
		\end{flushright}

	\begin{pro}
	\label{Analyticity}
		There exists a unique solution \( q_n(p) \in \ell_2(\mathbb{Z}) \) of (\ref{Eqr1}) and it is analytic on the imaginary axis for \( p \neq 0, \bar{p} \).
	\end{pro}
	
	\emph{Proof:} 
		The key point will be the application of the analytic Fredholm theorem to the operator \( \mathcal{L}(p) \) (Theorem VI.14 of \cite{Reed1}), in order to prove that \( (I - \mathcal{L}(p))^{-1} \) exists for \( p \neq 0, \bar{p} \). 
		\newline
		Since there is no non-zero solution in \( L^2_{\mathrm{loc}}(\mathbb{R}^+) \) of the homogeneous equation associated to (\ref{Equation}) (see the Proposition \ref{Homogeneous}), then the homogeneous equation associated to (\ref{Eqr1}) has only the trivial solution in \( \ell_2(\mathbb{Z}) \). Moreover the operator \( \mathcal{L} \) is compact and thus analytic Fredholm theorem applies. The result easily follows, because \( g(p) \in \ell_2(\mathbb{Z}) \) and each \( g_n(p) \) is analytic for \( p \neq 0, \bar{p} \).
		\begin{flushright} 
			\( \Box \)
		\end{flushright}

We can now study the equation (\ref{Eqr1}) in a neighborhood of \( \bar{p} \) (if \( \bar{p} \neq 0 \)). An important preliminary result is the following 

	\begin{lem}
	\label{r_n}
		Let (\ref{Conditions2}) and the genericity condition (\ref{Genericity}) be satisfied by \( \{ \alpha_n \} \). The system of equations 
		\beq
		\label{Eqrn2}
			r_n =  - \frac{4 \pi}{4 \pi \alpha_0 + \sqrt{\omega n - ip}} \bigg\{ \: \underset{k \neq n,\bar{n}}{\sum_{k \in \mathbb{Z}}} \alpha_{k-n} r_k + h_n(p) \bigg\}
		\eeq
		has a unique solution \( \{ r_n \} \in \ell_2(\mathbb{Z} \setminus \{\bar{n}\}) \) in a pure imaginary neighborhood of \( \bar{p} \), where \( \bar{n} \in \mathbb{Z} \) and \( \bar{p} \in \mathcal{I} \), \( \Re(\bar{p}) = 0 \), are defined by (\ref{Integer}), for every \( h_n(p) \) such that
		\bdm	
			h_n^{\prime}(p) \equiv \frac{h_n(p)}{4 \pi \alpha_0 + \sqrt{\omega n - ip}}
		\edm
		belongs to \( \ell_2(\mathbb{Z} \setminus \{ \bar{n} \}) \).
		\newline
		Moreover, if \( h_n(p) \) is analytic in a neighborhood of \( \bar{p} \), the solution is analytic in the same neighborhood.
	\end{lem}

	\emph{Proof:}
		Equation (\ref{Eqrn2}) is of the form
		\bdm
			r = \mathcal{L}^{\prime} r + h^{\prime}
		\edm
		where \( h^{\prime} \equiv \{ h_n^{\prime} \} \) belongs to \( \ell_2(\mathbb{Z} \setminus \{\bar{n}\}) \) and \( \mathcal{L}^{\prime} \) is a compact operator (see Proposition \ref{Compact}).
		\newline 
		In order to apply analytic Fredholm theorem to the operator \( \mathcal{L}^{\prime} \), we need to prove that there is no non-zero solution in a neighborhood of \( \bar{p} \) of the homogeneous equation. Suppose that the contrary is true, so that \( \{ R_n \} \in \ell_2(\mathbb{Z} \setminus \{\bar{n}\}) \) is a non-zero solution of
		\bdm
			R_n =  - \frac{4 \pi}{4 \pi \alpha_0 + \sqrt{\omega n - ip}} \underset{k \neq n,\bar{n}}{\sum_{k \in \mathbb{Z}}} \alpha_{k-n} R_{n}
		\edm
		Multiplying both sides of equation above by \( R_n^* \) and summing over \( n \in \mathbb{Z} \setminus \{\bar{n}\} \), one has
		\bdm
			\underset{n \neq \bar{n}}{\sum_{n \in \mathbb{Z}}} \sqrt{\omega n -ip} \:\: \big| R_n \big|^2 = - 4 \pi \underset{n,k \neq \bar{n}}{\sum_{n,k \in \mathbb{Z}}} {R_n}^* \alpha_{k-n}  R_{k}
		\edm
		and, since the right hand side is real, 
		\bdm
			\Im \bigg[ \: \underset{n \neq \bar{n}}{\sum_{n \in \mathbb{Z}}} \sqrt{\omega n -ip} \:\: \big| R_n \big|^2 \: \bigg] = 0
		\edm
		for \( p = i \lambda \), \( 0 < \lambda < \omega \), and then \( R_n = 0 \) for \( n < 0 \). Now suppose that \( R \neq 0 \) and let \( n_0 \in \mathbb{N} \) be such that \( R_n = 0 \), \( n < n_0 \), and \( R_{n_0} \neq 0 \) (hence \( n_0 \geq 0 \)). Fixing \( R_{\bar{n}} = 0 \), for each \( n < n_0 \) the homogeneous equation gives
		\bdm
			\sum_{k = n_0}^{\infty} \alpha_{k-n} R_{k} = 0
		\edm
		or, setting \( k = n_0 -1 + k^{\prime} \), for \( n \geq 0 \),
		\bdm
			\sum_{k^{\prime} = 1}^{\infty} \alpha_{k^{\prime} + n} R_{n_0-1+k^{\prime}} = 0
		\edm
		which implies (see (\ref{Conditions2})), for each \( n \geq 0 \),
		\bdm
			\Big(  R^{\prime} \: , \mathcal{T}^n \alpha \Big)_{\ell_2(\mathbb{N})} = 0
		\edm
		where \( R^{\prime}_n = R^*_{n_0-1+n} \) and \( ( \cdot \: , \: \cdot ) \) stands for the standard scalar product on \( \ell_2(\mathbb{N}) \).
		\newline
		If \( \{ \alpha_n \} \) satisfies the genericity condition (\ref{Genericity}), \( R^{\prime} \) has to be orthogonal also to \( e_1 \) and then \( R_{n_0} = 0 \), which is a contradiction. Therefore \( R = 0 \). 
		\newline
		The first part of the Lemma then follows from analyticity of \( \mathcal{L}^{\prime}(p) \) and analytic Fredholm theorem. Moreover if \( \{ h_n(p) \} \) is analytic in a neighborhood of \( \bar{p} \), analyticity of the solution is a straightforward consequence.
		\begin{flushright} 
			\( \Box \)
		\end{flushright}
 
	\begin{pro}
	\label{Poles}
		If \( \{ \alpha_n \} \) satisfies (\ref{Conditions2}) and the genericity condition with respect to \( \mathcal{T} \) (\ref{Genericity}), the unique solution \( \{ q_n \} \in \ell_2(\mathbb{Z}) \) of (\ref{Eqr1}) in analytic on the imaginary axis except at most for \( p = 0 \).
	\end{pro}

	\emph{Proof:}
		If \( (4 \pi \alpha_0)^2 = N \omega \) for some \( N \in \mathbb{N} \) (resonant case) there is nothing to prove, since the coefficients of (\ref{Eqr1}) fails to be analytic only at \( p = 0 \). On the other hand, in the non resonant case, Proposition \ref{Analyticity} guarantees analyticity on imaginary axis for \( p \neq 0,\bar{p} \). Therefore it is sufficient to study the behavior of the solution in a neighborhood of \( \bar{p} \), where the coefficients of (\ref{Eqr1}) have a singularity. We are going to prove that in fact the solution is analytic at \( \bar{p} \).
		\newline
		The strategy of the proof is to analyze separately the terms \( q_n \), \( n \neq \bar{n} \), \( \bar{n} \) being defined in (\ref{Integer}), and then prove that also \( q_{\bar{n}} \) is analytic in a neighborhood of \( \bar{p} \).
		\newline 
		By Lemma \ref{r_n} there is a unique solution of the system
		\beq
		\label{Eqtn}
			t_n = - \frac{4 \pi}{4 \pi \alpha_0 + \sqrt{\omega n - ip}} \underset{k \neq n,\bar{n}}{\sum_{k \in \mathbb{Z}}} \alpha_{k-n}  t_{k} - \frac{4 \pi \alpha_{\bar{n}-n}}{4 \pi \alpha_0 + \sqrt{\omega n - ip}}
		\eeq
		Setting \( q_n = r_n + t_n q_{\bar{n}} \), \( n \neq \bar{n} \), on (\ref{Eqr1}), one has
		\bdm
			r_n + t_n q_{\bar{n}}  = - \frac{4 \pi}{4 \pi \alpha_0 + \sqrt{\omega n - ip}} \bigg\{ \alpha_{\bar{n}-n} q_{\bar{n}} + \underset{k \neq n,\bar{n}}{\sum_{k \in \mathbb{Z}}} \alpha_{k-n} \big( r_k + t_k q_{\bar{n}} \big) \bigg\} +
		\edm
		\bdm
			- \frac{2 i \sqrt{2\pi}}{4 \pi \alpha_0 + \sqrt{\omega n - ip}} \frac{1 - \sqrt{\omega n -ip}}{1 + ip - \omega n} 
		\edm
		and therefore the equation for \( \{ r_n \} \), \( n \neq \bar{n} \), becomes
		\beq
		\label{Eqrn}
			r_n =  - \frac{4 \pi}{4 \pi \alpha_0 + \sqrt{\omega n - ip}} \bigg\{ \: \underset{k \neq 0,-n}{\sum_{k \in \mathbb{Z}}} \alpha_k r_{n+k} + \frac{i}{\sqrt{2 \pi}} \: \frac{1 - \sqrt{\omega n -ip}}{1 + ip - \omega n} \bigg\} 
		\eeq
		while \( q_{\bar{n}} \) satisfies the equation
		\bdm	
			q_{\bar{n}} = - \frac{4 \pi}{4 \pi \alpha_0 + \sqrt{\omega \bar{n}-ip}} \bigg\{ \: \underset{k \neq \bar{n}}{\sum_{k \in \mathbb{Z}}} \alpha_{k-\bar{n}} \big( r_k + t_k q_{\bar{n}} \big) + \frac{i}{\sqrt{2 \pi}} \: \frac{1 - \sqrt{\omega \bar{n} -ip}}{1 + ip - \omega \bar{n}} \bigg\}
		\edm
		or
		\bdm
			\bigg[ 4 \pi \alpha_0 + \sqrt{\omega \bar{n}-ip} + 4 \pi \: \underset{k \neq \bar{n}}{\sum_{k \in \mathbb{Z}}} \alpha_{k-\bar{n}} t_k \bigg] \: q_{\bar{n}} =  - 4 \pi \underset{k \neq \bar{n}}{\sum_{k \in \mathbb{Z}}} \alpha_{k-\bar{n}} r_{k} - \frac{2i \sqrt{2\pi}}{1 + \sqrt{\omega \bar{n} -ip}} 
		\edm
		Since the last term is analytic in a neighborhood of \( \bar{p} \) and \( \{ t_n \} \), \( \{ r_n \} \in \ell_2(\mathbb{Z} \setminus \{ \bar{n} \}) \) are both analytic, as it follows applying Lemma \ref{r_n} above to (\ref{Eqtn}) and (\ref{Eqrn}), it is sufficient to prove that
		\bdm
			\underset{k \neq \bar{n}}{\sum_{k \in \mathbb{Z}}} \: \alpha_{k-\bar{n}} \tilde{t}_k \neq 0
		\edm
		where
		\bdm
			\tilde{t}_n \equiv t_n(p)\big|_{p = \bar{p}} 
		\edm 
		Assume that the contrary is true: from equation (\ref{Eqtn}) we obtain
		\bdm
			\underset{n \neq \bar{n}}{\sum_{n \in \mathbb{Z}}} \big( 4 \pi \alpha_0 + \sqrt{\omega n - i\bar{p}} \big) \: \big| \tilde{t}_n \big|^2 =  - 4 \pi \underset{n,k \neq \bar{n}, n \neq k}{\sum_{n,k \in \mathbb{Z}}} {\tilde{t}_n}^* \alpha_{k-n}  \tilde{t}_{k} - 4 \pi \underset{n \neq \bar{n}}{\sum_{n \in \mathbb{Z}}} \: \alpha^*_{n-\bar{n}} {\tilde{t}_n}^* = 
		\edm
		\bdm
			= - 4 \pi \underset{n,k \neq \bar{n}, n \neq k}{\sum_{n,k \in \mathbb{Z}}} {\tilde{t}_n}^* \alpha_{k-n}  \tilde{t}_{k}
		\edm
		where we have used condition 2 in (\ref{Conditions2}). The previous equation implies (the right hand side is real) \( \tilde{t}_n = 0 \), \( \forall n < \bar{N} = \frac{i \bar{p}}{\omega} \) and then, since \( -1 < \bar{N} < 0 \),   \( \tilde{t}_n = 0 \), \( \forall n < 0 \). Hence from (\ref{Eqtn}) we have, \( \forall n < 0 \),
		\bdm
			\underset{k \neq \bar{n}}{\sum_{k \geq 0}} \alpha_{k-n} \tilde{t}_{k} + \alpha_{\bar{n}-n} = 0
		\edm
		Now supposing without loss of generality that \( \tilde{t}_0 \neq 0 \) and setting \( T_n = \tilde{t}_{n-1} \), \( n \neq \bar{n}+1 \), and \( T_{\bar{n}+1} = 1 \), we obtain, \( \forall n \geq 0 \),
		\bdm
			\sum_{k = 1}^{\infty} \alpha_{k+n}  T_{k} = 0 
		\edm
		and using the genericity condition (\ref{Genericity}) (as in the proof of Lemma \ref{r_n}) we get \( T_1 = t_0 = 0 \), which is a contradiction. 
		\newline
		In conclusion \( q_{\bar{n}}\) is analytic in a neighborhood of \( \bar{p} \): analyticity of \( q_n \), \( n \neq \bar{n} \) is then a straightforward consequence of analyticity of \( \{ r_n \} \), \( \{ t_n \} \) and decomposition \( q_n = r_n + t_n q_{\bar{n}} \). The proof is then completed, since \( r_n \) and \( t_n \) belong to \( \ell_2(\mathbb{Z} \setminus \{ \bar{n} \}) \) in a neighborhood of \( p = \bar{p} \).
		\begin{flushright} 
			\( \Box \)
		\end{flushright}

\subsection{Behavior at \( p = 0 \)}

We shall now study the behavior of the solution of (\ref{Eqr1}) on the imaginary axis at the origin. With the choice (\ref{Branch}) for the branch cut of the square root, it is clear that we must expect branch points of \( \tilde{q}(p) \), solution of (\ref{Eq1}), at \( p = i\omega n \), \( n \in \mathbb{Z} \), which should imply a branch point at \( p = 0 \) for each \( q_n \) in (\ref{Eqr1}).
\newline
We are going to show that \( q_n \), \( n \in \mathbb{Z} \) has a branch point at \( p = 0 \). The non-resonant case and the resonant one will be treated separately.

	\begin{pro}[non-resonant case]	
	\label{BranchPoints} 
		\mbox{}	\\
		If \( (4 \pi \alpha_0)^2 \neq N \omega \), \( \forall N \in \mathbb{N} \) and \( \{ \alpha_n \} \) satisfies (\ref{Conditions2}) and (\ref{Genericity}) (genericity condition), the solution of equation (\ref{Eqr1}) has the form \( q_n(p) = c_n(p) + d_n(p) \sqrt{p} \), \( n \in \mathbb{Z} \), in an imaginary neighborhood of \( p = 0 \), where the functions \( c_n(p) \) and \( d_n(p) \) are analytic at \( p = 0 \).
	\end{pro}

	\emph{Proof:}
		Setting \( q_n = r_n + t_n q_0 \), \( n \neq 0 \) and choosing a solution \( \{ t_n \} \in \ell_2(\mathbb{Z} \setminus \{ 0 \}) \) of the system of equations (\ref{Eqtn}) with \( \bar{n} = 0 \), we obtain that \( \{ r_n \} \) must satisfy (\ref{Eqrn}). It is easy to see that the result of Lemma \ref{r_n} holds also in a neighborhood of \( \bar{p} = 0 \) with \( \bar{n} = 0 \), so that \( \{ r_n \} \), \( \{ t_n \} \in \ell_2(\mathbb{Z} \setminus \{0\}) \) are unique and analytic at \( p = 0 \).
		\newline
		Thus it is sufficient to prove that \( q_0 \), which is solution of 
		\bdm
			\bigg[ 4 \pi \alpha_0 + \sqrt{-ip} + 4 \pi \: \underset{k \neq 0}{\sum_{k \in \mathbb{Z}}} \alpha_k t_k \bigg] \: q_0 =  - 4 \pi \underset{k \neq 0}{\sum_{k \in \mathbb{Z}}} \alpha_k r_{k} - \frac{2i \sqrt{2\pi} (1 - \sqrt{-ip})}{1 + ip} 
		\edm
		has the required behavior near \( p = 0 \).
		\newline
		First, setting \( t_n^0 = t_n(p = 0) \), we have to prove that
		\bdm
			\underset{k \neq 0}{\sum_{k \in \mathbb{Z}}} \alpha_k t^0_k \neq - \alpha_0
		\edm
		but, assuming that the contrary is true and  multiplying both sides of equation (\ref{Eqtn}) by \( {t_n^0}^* \) and summing over \( n \in \mathbb{Z} \), \( n \neq 0 \), one has
		\bdm
			 \sum_{n \in \mathbb{Z}} \sqrt{\omega n} \: |t^0_n|^2 = - 4 \pi \underset{n,k \neq 0}{\sum_{n,k \in \mathbb{Z}}} {t_n^0}^* \alpha_{k-n} t^0_{k} + 4 \pi \alpha_0
		\edm
		and then, because of genericity condition (\ref{Genericity}), \( \{ t^0_n \} = 0 \), \( \forall n \in \mathbb{Z} \setminus \{ 0 \} \), which is impossible, since \( \{ t_n \} \) solves (\ref{Eqtn}).
		\newline
		Now, calling
		\bdm
			F \equiv 4 \pi \underset{k \neq 0}{\sum_{k \in \mathbb{Z}}} \alpha_k t_k 
		\edm
		and 
		\bdm
			G \equiv - 4 \pi \underset{k \neq 0}{\sum_{k \in \mathbb{Z}}} \alpha_k r_k
		\edm
		we have
		\bdm
			\bigg[ 4 \pi \alpha_0 + \sqrt{-ip} + F \bigg] \: q_0 = G + \frac{2i \sqrt{2 \pi}(1- \sqrt{-ip})}{1+ip}
		\edm
		and
		\bdm	
			 q_0 = F^{\prime} + \sqrt{p} \:\: G^{\prime}
		\edm
		where \( F^{\prime} \) is analytic in a neighborhood of \( p = 0 \), because of analyticity of \( F \) and \( G \), and 
		\beq
		\label{SquareRoot}
			G^{\prime} \equiv - \frac{2i \sqrt{-2 \pi i} \: ( 4 \pi \alpha_0 + F + 1 ) + \sqrt{-i} \: (1+ip) \: G}{(1+ip)[(4 \pi \alpha_0 + F)^2+ip]}
		\eeq
		\begin{flushright} 
			\( \Box \)
		\end{flushright}

The resonant case, i.e. \( 4 \pi \alpha_0 = - \sqrt{\omega N} \) for some \( N \in \mathbb{N} \), is not so different from the non-resonant one and we shall prove that the solution has the same behavior at the origin. The proof is slightly different because we need to show the absence of a pole at \( p = 0 \): from (\ref{Eqr1}) one has
	\bdm
		q_N(p) = \frac{4 \pi}{ \sqrt{\omega N} - \sqrt{\omega N - ip}} \bigg\{ \: \underset{k \neq 0}{\sum_{k \in \mathbb{Z}}} \: \alpha_k \: q_{n+k}(p) + \frac{i}{\sqrt{2 \pi}} \frac{1 - \sqrt{\omega N - ip }}{1 + ip - \omega N} \bigg\}
	\edm
and the coefficients have a singularity at \( p = 0 \).
\newline
We are going to prove that in fact the solution has no pole at the origin: proceeding as in the proof of Proposition \ref{Poles}, let us begin with a preliminary result, which take the place of Lemma \ref{r_n}:

	\begin{lem}
	\label{r_n2}
		Let (\ref{Conditions2}) and the genericity condition (\ref{Genericity}) be satisfied by \( \{ \alpha_n \} \). The system of equations 
		\beq
		\label{Eqrn3}
			r_n =  \frac{4 \pi}{\sqrt{\omega N} - \sqrt{\omega n - ip}} \bigg\{ \: \underset{k \neq 0,-n}{\sum_{k \in \mathbb{Z}}} \alpha_k r_{n+k} + h_n(p) \bigg\}
		\eeq
		has a unique solution \( \{ r_n \} \in \ell_2(\mathbb{Z} \setminus \{N\}) \) in a pure imaginary neighborhood of \( p = 0 \), for every \( h_n(p) \) such that
		\bdm	
			h_n^{\prime}(p) \equiv \frac{h_n(p)}{\sqrt{\omega N} - \sqrt{\omega n - ip}}
		\edm
		belongs to \( \ell_2(\mathbb{Z} \setminus \{ N \}) \).
		\newline
		Moreover, if \( h_n(p) \) is analytic in a neighborhood of \( p = 0 \), the solution is analytic in the same neighborhood.
	\end{lem}
 
	\emph{Proof:}
		We shall proceed as in the proof of Proposition \ref{Poles}, separating the contribution of \( r_N \), which may be singular: setting \( r_n = u_n + v_n r_N \), \( n \neq 0 , N \), on (\ref{Eqrn3}), one has
		\bdm
			u_n + v_n r_N  = \frac{4 \pi}{\sqrt{\omega N} - \sqrt{\omega n - ip}} \bigg\{ \alpha_{N-n} r_N + \underset{k \neq 0,-n,N-n}{\sum_{k \in \mathbb{Z}}} \alpha_k \big( u_{n+k} + v_{n+k} r_N \big) \bigg\} +
		\edm
		\bdm
			+ \frac{2i \sqrt{2\pi}}{\sqrt{\omega N} - \sqrt{\omega n - ip}} \frac{1 - \sqrt{\omega n -ip}}{1 + ip - \omega n} 
		\edm
		and requiring that \( \{ v_n \} \), \( n \neq 0, N \), solves
		\beq
		\label{Eqvn}
			v_n = \frac{4 \pi}{\sqrt{\omega N} - \sqrt{\omega n - ip}} \underset{k \neq 0,-n, N-n}{\sum_{k \in \mathbb{Z}}} \alpha_k  v_{n+k} + \frac{4 \pi \alpha_{N-n}}{\sqrt{\omega N} - \sqrt{\omega n - ip}}
		\eeq
		the equation for \( \{ u_n \} \), \( n \neq 0, N \), becomes
		\beq
		\label{Equn}
			u_n =  \frac{4 \pi}{\sqrt{\omega N} - \sqrt{\omega n - ip}}  \bigg\{ \: \underset{k \neq 0,-n, N-n}{\sum_{k \in \mathbb{Z}}} \alpha_k u_{n+k} + \frac{i}{\sqrt{2 \pi}} \: \frac{1 - \sqrt{\omega n -ip}}{1 + ip - \omega n} \bigg\} 
		\eeq
		Moreover \( r_N \) satisfies 
		\bdm	
			r_N = \frac{4 \pi}{\sqrt{\omega N} - \sqrt{\omega N - ip}} \bigg\{ \: \underset{k \neq 0, -N}{\sum_{k \in \mathbb{Z}}} \alpha_k \big( u_{k} + v_k r_N \big) + \frac{i}{\sqrt{2 \pi}} \: \frac{1 - \sqrt{\omega n -ip}}{1 + ip - \omega n} \bigg\}
		\edm
		or
		\bdm
			\bigg[ \sqrt{\omega N} - \sqrt{\omega N -ip} - 4 \pi \: \underset{k \neq 0, N}{\sum_{k \in \mathbb{Z}}} \alpha_{k-N} v_k \bigg] \: r_N =  
		\edm
		\bdm
			= 4 \pi \underset{k \neq 0, N}{\sum_{k \in \mathbb{Z}}} \alpha_{k-N} u_{k} + \frac{i}{\sqrt{2 \pi}} \: \frac{1 - \sqrt{\omega n -ip}}{1 + ip - \omega n}
		\edm
		Applying the discussion contained in the proof of Lemma \ref{r_n}, it is not difficult to see that the solutions of equations (\ref{Equn}) and (\ref{Eqvn}) are analytic in a neighborhood of the origin and belong to \( \ell_2(\mathbb{Z} \setminus \{ 0,N \}) \). Therefore it remains to prove that (setting \( v^0_n = v_n(p=0) \))
		\bdm
			\underset{k \neq 0, N}{\sum_{k \in \mathbb{Z}}} \alpha_{k-N} v^0_k \neq 0
		\edm
		but the argument in the proof of Proposition \ref{Poles} excludes this possibility, if \( \{ \alpha_n \} \) satisfies the genericity condition. The proof is then completed, because analyticity of \( r_N \) implies analyticity of all \( r_n \), \( n \neq 0, N \). 
		\begin{flushright} 
			\( \Box \)
		\end{flushright}

	\begin{pro}[resonant case]	
	\label{BranchPointsR} 
		\mbox{}	\\
		If \( (4 \pi \alpha_0)^2 = N \omega \), for some \( N \in \mathbb{N} \) and \( \{ \alpha_n \} \) satisfies (\ref{Conditions2}) and (\ref{Genericity}) (genericity condition), the solution of equation (\ref{Eqr1}) has the form \( q_n(p) = c_n(p) + d_n(p) \sqrt{p} \), \( n \in \mathbb{Z} \), in an imaginary neighborhood of \( p = 0 \), where the functions \( c_n(p) \) and \( d_n(p) \) are analytic at \( p = 0 \).
	\end{pro}

	\emph{Proof:}	
		See the proof of Proposition \ref{BranchPoints} and Lemma \ref{r_n2} above.
		\begin{flushright} 
			\( \Box \)
		\end{flushright}

\subsection{Complete ionization in the generic case}

Summing up the results about the behavior of the Laplace transform \( \tilde{q}(p) \) of \( q(t) \) we can state the following
	
	\begin{teo}
	\label{Decay}
		If \( \{ \alpha_n \} \) satisfies (\ref{Conditions2}) and the genericity condition (\ref{Genericity}) with respect to \( \mathcal{T} \), as \( t \rightarrow \infty \),
		\beq
			|q(t)| \leq A \:  t^{-\frac{3}{2}} + R(t)
		\eeq
		where \( A \in \mathbb{R} \) and \( R(t) \) has an exponential decay, \( R(t) \sim C e^{-Bt} \) for some \( B > 0 \).
	\end{teo}

	\emph{Proof:}	Propositions \ref{Analyticity}, \ref{Poles} and \ref{BranchPoints} guarantee that \( \tilde{q}(p) \) is analytic on the closed right half plane, except branch point singularities on the imaginary axis at \( p = i \omega n \), \( n \in \mathbb{Z} \).
		\newline
		Therefore we can chose a integration path for the inverse of Laplace transform of \( \tilde{q}(q) \) along the imaginary axis like in \cite{Cost1}. 
		\newline	
		Proposition \ref{BranchPoints} implies that the contribution of the branch point at \( p = 0 \) is given by the integral
		\bdm
			2 i \int_0^{\infty} dp \: \sqrt{p} \:\: G^{\prime}(-p) \: e^{-pt} 
		\edm
		where \( G^{\prime} \), defined in (\ref{SquareRoot}), is a bounded analytic function on the negative real line: from explicit expression of \( F \) and \( G \) and equations (\ref{Eqrn}) and (\ref{Eqtn}), it is clear that \( G^{\prime} \) is analytic and \( \lim_{p \rightarrow \infty} G^{\prime}(-p) = 0 \) on the real line. So that the corresponding asymptotic behavior as \( t \rightarrow \infty \) is 
		\bdm
			\bigg| \int_0^{\infty} dp \: \sqrt{p} \:\: G^{\prime}(-p) \: e^{-pt} \bigg| \leq C \int_0^{\infty} dp \: \sqrt{p} \:\: e^{-pt} = A \: t^{-\frac{3}{2}}
		\edm
		Let us consider now the contribution of branch points at \( p = i \omega n \), \( n \neq 0 \): from Propositions \ref{BranchPoints} and \ref{BranchPointsR} it follows that, in a neighborhood of \( p = 0 \), 
		\bdm
			q_n(p) = c_n(p) + d_n(p) \: \sqrt{p}
		\edm 
		where \( c_n(p) \) and  \( d_n(p) \) are analytic at \( p = 0 \). Moreover using the decomposition \( q_n = r_n + t_n q_0 \), \( n \neq 0 \), as in the proof of Proposition \ref{BranchPoints} and \ref{BranchPointsR}, and studying the equation (\ref{Eqtn}) for \( t_n \), we immediately obtain \( \{ d_n \} \in \ell_1(\mathbb{Z} \setminus \{ 0 \}) \), because of condition 2 in (\ref{Conditions2}). Since \( q_n(p) = \tilde{q}(p + i \omega n) \), the contribution of singularities at \( p = i \omega n \), \( n \neq 0 \), is then given by
		\bdm
			2 \underset{n \neq 0}{\sum_{n \in \mathbb{Z}}} \int_{i \omega n-\infty}^{i \omega n} dp \:\: d_n(p-i\omega n) \sqrt{p-i \omega n} \:\: e^{pt} =
		\edm
		\bdm
			= 2i \int_{0}^{\infty} dp \: \bigg\{ \underset{n \neq 0}{\sum_{n \in \mathbb{Z}}} \: d_n(-p) \: e^{i \omega n t} \bigg\} \sqrt{p} \:\: e^{-pt} =
		\edm
		and the series
		\bdm
			\underset{n \neq 0}{\sum_{n \in \mathbb{Z}}} \: d_n(-p) \: e^{i \omega n t}
		\edm
		converges uniformly to a bounded function of \( t \), because \( \{ d_n \} \in \ell_1(\mathbb{Z} \setminus \{ 0 \}) \).
		\newline
		Adding up the contributions of every branch cut, one obtain the required leading term in the asymptotic behavior. Indeed the rest function \( R(t) \) is given by the contribution of poles outside the imaginary axis and then shows an exponential decay as \( t \rightarrow \infty \).
		\begin{flushright} 
			\( \Box \)
		\end{flushright}

A straightforward consequence of Theorem \ref{Decay} is that the scalar product (and thus the survival probability of the bound state)
	\bdm
		\theta(t) = \Big( \varphi_{\alpha(0)} \: , \Psi_t \Big)_{L^2(\mathbb{R}^3)} 
	\edm
tends to 0 when \( t \rightarrow \infty \):

	\begin{cor}
		\label{Survive}
		If \( \{ \alpha_n \} \) satisfies (\ref{Conditions2}) and the genericity condition (\ref{Genericity}) with respect to \( \mathcal{T} \), the system shows asymptotic complete ionization and, as \( t \rightarrow \infty \), 
		\bdm
			| \theta(t) | \leq D \: t^{-\frac{3}{2}} + E(t)
		\edm
		where \( D \in \mathbb{R} \) and \( E(t) \) has an exponential decay.
	\end{cor}
	
	\emph{Proof:}
		The Laplace transform of \( \theta(t) \) can be expressed in the following way (see the proof of Proposition \ref{Analy})
		\bdm
			\tilde{\theta}(p) = \tilde{Z}_1(p) + \tilde{Z}_2(p) \: \tilde{q}(p)
		\edm 
		where \( \tilde{Z}_1(p) \) is analytic on the closed right half plane and \( \tilde{Z}_2(p) \) has only a branch point at the origin of the form \( a_1 + a_2 \sqrt{p} \).
		\newline
		Hence \( \tilde{\theta}(p) \) has the same singularities as \( \tilde{q}(p) \) and then its asymptotic behavior coincides with that of \( q(t) \), i.e.
		\bdm
			|\theta(t)| \leq D \: t^{-\frac{3}{2}} + E(t)
		\edm
		for some constant \( D \in \mathbb{R} \) and for a bounded function \( E(t) \) with exponential decay. 
		\begin{flushright} 
			\( \Box \)
		\end{flushright}

In the following we shall prove a stronger result about complete ionization of the system, namely that every state \( \Psi \in L^2(\mathbb{R}^3) \) is a scattering state\footnote{For the definition of scattering states of a time-dependent operator see e.g. \cite{Enss1,Howl1}.} for the operator \( H_{\alpha(t)} \), i.e. 
	\beq
	\label{Ioni}
		\lim_{t \rightarrow \infty} \frac{1}{t} \int_0^t d\tau \: \big\| F(|\vec{x}| \leq R) U(\tau,0) \Psi \big\|^2 = 0
	\eeq
where \( F(S) \) is the multiplication operator by the characteristic function of the set \( S \subset \mathbb{R}^3 \) and \( U(t,s) \) the unitary two-parameters family associated to \( H_{\alpha(t)} \) (see (\ref{Schro})).
\newline
In order to prove (\ref{Ioni}), we first need to study the evolution of a generic initial datum in a suitable dense subset of \( L^2(\mathbb{R}^3) \) and then we shall extend the result to every state using the unitarity of the evolution defined by (\ref{Schro}) (see e.g. \cite{Dell1}).
	
	\begin{pro}
	\label{DecayG}
		Let \( \Psi \in C^{\infty}_0(\mathbb{R}^3 \setminus \{ 0 \}) \) a smooth function with compact support away from \( 0 \) and \( q(t) \) be the solution of equation (\ref{Equation}) with initial condition \( \Psi_0 = \Psi \). If \( \{ \alpha_n \} \) satisfies (\ref{Conditions2}) and the genericity condition (\ref{Genericity}) with respect to \( \mathcal{T} \), as \( t \rightarrow \infty \),
		\beq
			| q(t) | \leq A \:  t^{-\frac{3}{2}} + R(t)
		\eeq
		where \( A \in \mathbb{R} \) and \( R(t) \) has an exponential decay, \( R(t) \sim C e^{-Bt} \) for some \( B > 0 \).
	\end{pro}

	\emph{Proof:}	
		The proof of Proposition \ref{Analy} still applies, considering 
		\bdm
			\theta^{\prime}(t) \equiv \Big( \Psi \: , \Psi_t \Big)_{L^2(\mathbb{R}^3)} 
		\edm
instead of \( \theta(t) \), so that \( \tilde{q}(p) \), solution of (\ref{Laplace}) with initial condition \( \Psi_0 = \Psi \), is analytic \( \forall p \) with \( \Re(p) > 0 \).
		\newline
		Hence we can consider the Laplace transform of equation (\ref{Equation}), which has the form (\ref{Laplace}) with
		\bdm
			f(p) = \sqrt{\frac{2}{\pi}} \: \sqrt{\frac{i}{p}} \: \int_0^{\infty} dt \: e^{-pt} \int_{\mathbb{R}^3} d^3 \vec{k} \:\: \hat{\Psi}(\vec{k}) \: e^{-ik^2t}
		\edm
		where \( \hat{\Psi}(\vec{k}) \) is the Fourier transform of \( \Psi(\vec{x}) \). 
		\newline
		The equation for \( \tilde{q}(p) \) is then given by
		\bdm
			\tilde{q}(p) =  - \frac{4 \pi}{4 \pi \alpha_0 + \sqrt{-ip}} \: \underset{k \neq 0}{\sum_{k \in \mathbb{Z}}} \: \alpha_k \: \tilde{q}(p+i \omega k) + \frac{g(p)}{4 \pi \alpha_0 + \sqrt{-ip}} 
		\edm
		where
		\bdm
			g(p) = \sqrt{\frac{2}{\pi}} \: \int_0^{\infty} dt \: e^{-pt} \int_{\mathbb{R}^3} d^3 \vec{k} \:\: \hat{\Psi}(\vec{k}) \: e^{-ik^2t}
		\edm
		It is now sufficient to show that the solution \( \tilde{q}(p) \) is also analytic on the imaginary axis except at most square root branch points at \( p = i \omega n \) as in the discussion of section 3.2 and 3.3.
		\newline
		For every smooth function \( \Psi \) with compact support, \( \hat{\Psi}(\vec{k}) \) is a smooth function with an exponential decay as \( k \rightarrow \infty \), so that
		\bdm
			g(is) = \lim_{r \rightarrow 0^+} \sqrt{\frac{2}{\pi}} \: \int_{\mathbb{R}^3} d^3 \vec{k} \:\: \frac{\hat{\Psi}(\vec{k})}{ r + (s + k^2)i} = - i \sqrt{\frac{2}{\pi}} \: \int_{\mathbb{R}^3} d^3 \vec{k} \:\: \frac{\hat{\Psi}(\vec{k})}{s + k^2}
		\edm
		is a bounded function for \( s > 0 \). Hence the function \( g(p) \) has no pole for \( \Im(p) \in (0 , \omega) \) and therefore the result contained in Proposition \ref{Poles} still holds.
		\newline
		Moreover
		\bdm
			g(0) = \sqrt{\frac{2}{\pi}} \int_{\mathbb{R}^3} d^3 \vec{k} \:\: \hat{\Psi}(\vec{k}) \int_0^{\infty} dt \: e^{-ik^2t} = -i \sqrt{\frac{2}{\pi}} \int_{\mathbb{R}^3} d^3 \vec{k} \:\: \frac{\hat{\Psi}(\vec{k})}{k^2}
		\edm
		which is again bounded, so that \( g(p) \) has at the origin at most a branch point singularity of the form \( a(p) + b(p) \sqrt{p} \): following the proofs of Proposition \ref{BranchPoints} and \ref{BranchPointsR}, we can show that \( \tilde{q}(p) \) has the same behavior at the origin.
		\newline
		In conclusion the solution is analytic on the closed right half plane except branch points at \( p = i \omega n \), \( n \in \mathbb{Z} \), of the form \( a(p) + b(p)\sqrt{p-i\omega n} \). The proof of Theorem \ref{Decay} then implies that \( q(t) \) has the prescribed behavior as \( t \rightarrow \infty \).
		\begin{flushright} 
			\( \Box \)
		\end{flushright}

	\begin{teo}
	\label{Scattering}
		If \( \{ \alpha_n \} \) satisfies (\ref{Conditions2}) and the genericity condition (\ref{Genericity}) with respect to \( \mathcal{T} \), every \( \Psi \in L^2(\mathbb{R}^3) \) is a scattering state of \( H_{\alpha(t)} \), i.e.
		\bdm
			\lim_{t \rightarrow \infty} \frac{1}{t} \int_0^t d\tau \: \big\| F(|\vec{x}| \leq R) U(\tau,0) \Psi \big\|^2 = 0
		\edm
	\end{teo}
		
	\emph{Proof:}
		We shall restrict the proof to the dense subset of \( L^2(\mathbb{R}^3) \) given by smooth functions with compact support and then we shall extend the result to every state using the unitarity of the evolution defined by (\ref{State}) (see e.g. \cite{Dell1}). Actually we are going to prove an equivalent but slightly different statement, i.e. \(  \forall \varepsilon > 0 \), there exists \( t_0 \) such that \( \forall t > t_0 \),
		\bdm
			\big\| F(|\vec{x}| \leq R) U(t,0) \Psi \big\| \leq \varepsilon
		\edm
		The evolution of an initial state \( \Psi \) according to (\ref{State}) is given by
		\beq
		\label{Decomposition1}
			\Psi_t(\vec{x}) = U(t,s)\Psi_s (\vec{x}) = U_0(t-s) \Psi_s (\vec{x}) + i \int_s^t d \tau \: q(\tau) \: U_0(t-\tau ; \vec{x})
		\eeq
		Moreover, since \( \Psi_t \in \mathcal{D}(H_{\alpha(t)}) \), the following decomposition holds
		\beq
		\label{Decomposition2}
			\Psi_t(\vec{x}) = \varphi_t(\vec{x}) + \frac{q(t)}{4 \pi |\vec{x}|}
		\eeq
		where \( q(t) \) is the solution of (\ref{Equation}), \( \varphi_t \in H^2_{\mathrm{loc}}(\mathbb{R}^3) \) and
		\bdm
			\varphi_t(0) = \alpha(t) q(t)
		\edm
		We are going to show that, if \( q(t) \in L^1(\mathbb{R}^+) \), \( \Psi_t \) satisfies the required property. Let us start analyzing the second term in (\ref{Decomposition1}): imposing the unitarity condition of the evolution we have
		\bdm
			\| \Psi_s \|^2 = \| \Psi_t \|^2 = \bigg\|  U_0(t-s) \Psi_s (\vec{x}) + i \int_s^t d \tau \: q(\tau) \: U_0(t-\tau ; \vec{x}) \bigg\|^2
		\edm
		and then
		\bdm
			\bigg\| \int_s^t d \tau \: q(\tau) \: U_0(t-\tau ; \vec{x}) \bigg\|^2 = 2 \Im \bigg( \int_s^t d \tau \: q(\tau) \: U_0(t-\tau ; \vec{x}) \: , \: U_0(t-s) \Psi_s (\vec{x}) \bigg) =
		\edm
		\bdm
			= 2 \Im \bigg[ \int_s^t d \tau \: q^*(\tau) \Big( e^{-iH_0(\tau -s)} \Psi_s \Big) (0) \bigg]
		\edm
		but, using the decomposition (\ref{Decomposition2}),
		\bdm
			\Big( e^{-iH_0(s-\tau)} \Psi_s \Big)(0) = \Big( e^{-iH_0(s-\tau)} \varphi_s \Big)(0) + \int_{\mathbb{R}^3} d^3 \vec{k} \: e^{-ik^2(\tau -s)} \: \frac{q(s)}{(2 \pi)^3 k^2} =   
		\edm
		\bdm
			= \Big( e^{-iH_0(s-\tau)} \varphi_s \Big)(0) + \frac{q(s)}{4 \pi \sqrt{\pi i} \sqrt{\tau-s}}
		\edm
		Since \( \varphi_s \in H^2_{\mathrm{loc}}(\mathbb{R}^3) \), the absolute value of the first term on the right hand side is bounded by a constant \( c(\tau,s) < \infty \) such that \( c(s,s) = q(s) \) and 
		\bdm
			\lim_{\tau \rightarrow \infty} c(\tau,s) = 0 
		\edm
		Hence there exists \( s_1(\varepsilon) > 0 \) such that, \(  \forall s > s_1 \), 
		\bdm
			2 \bigg| \int_s^t d \tau \: q^*(\tau) \Big( e^{-iH_0(s-\tau)} \varphi_s \Big)(0) \bigg| \leq \frac{2\varepsilon^2}{9}
		\edm
		if \( q(t) \in L^1(\mathbb{R}^+) \). Moreover by the same reason there exists \( s_2(\varepsilon) >0 \) such that \( \forall s>s_2 \),
		\bdm
			2 \bigg| \int_s^t d \tau \: q^*(\tau) \frac{q(s)}{4 \pi \sqrt{\pi i} \sqrt{\tau-s}} \bigg| \leq \frac{2\varepsilon^2}{9}
		\edm
		Setting \( s_0(\varepsilon) = \max(s_1(\varepsilon),s_2(\varepsilon)) \), one has \( \forall s > s_0 \)
		\beq
		\label{Ineq1}
			\bigg\| \int_s^t d \tau \: q(\tau) \: U_0(t-\tau ; \vec{x}) \bigg\| \leq \frac{2 \varepsilon}{3}
		\eeq
		so that the whole \(L^2-\)norm of the second term in decomposition (\ref{Decomposition1}) is suitably small for \( s > s_0 \).
		\newline
		On the other hand the first term in (\ref{Decomposition1}) is the free evolution of a \(L^2-\)function and hence there exists \( \delta(\varepsilon) > 0 \) such that \( \forall t > s+\delta \) and \( \forall R < \infty \),
		\beq
		\label{Ineq2}
			\big\| F(|\vec{x}| \leq R) U(t-s) \Psi_s \big\| \leq \frac{\varepsilon}{3}
		\eeq
		Setting \( t_0(\varepsilon) = s_0(\varepsilon) + \delta(\varepsilon) \), from (\ref{Decomposition1}), (\ref{Ineq1}) and (\ref{Ineq2}) one has
		\bdm
			\big\| F(|\vec{x}| \leq R) \Psi_t \big\| \leq \varepsilon
		\edm
		\( \forall t > t_0 \), if \( q(t) \in L^1(\mathbb{R}^+) \).
		\newline
		By Proposition \ref{DecayG} the inequality is then satisfied by every \( \Psi \in C_0^{\infty}(\mathbb{R}^3 \setminus \{0\}) \): unitarity of the family \( U(t,s) \) allows to extend the result to the whole Hilbert space \( L^2(\mathbb{R}^3) \).
		\begin{flushright} 
			\( \Box \)
		\end{flushright}

	\begin{cor}
		If \( \{ \alpha_n \} \) satisfies (\ref{Conditions2}) and the genericity condition with respect to \( \mathcal{T} \) (\ref{Genericity}), the discrete spectrum of the Floquet operator associated to \( H_{\alpha(t)} \),
		\bdm
			K \equiv -i \frac{\partial}{\partial t} + H_{\alpha(t)}
		\edm
		is empty.
	\end{cor}
		
	\emph{Proof:}
		The result is a straightforward consequence of Theorem \ref{Scattering}: every eigenvector of \( K \) differs from a periodic function by a phase factor and hence can not satisfy (\ref{Ioni}).
		\begin{flushright} 
			\( \Box \)
		\end{flushright}

\section{CASE II: \( \alpha_0 = 0 \)}

If \( \alpha(t) = \alpha_0 = 0 \) does not depend on time, the problem has a simple solution: the spectrum of \( H_{\alpha(t)} \) is absolutely continuous and equal to the positive real line, with a resonance at the origin; hence there is no bound state and the system shows complete ionization independently on the initial datum.
\newline
On the other hand if \( \alpha(t) \) is a zero mean function, we shall see that the genericity condition (\ref{Genericity}) is still needed to have complete ionization.
\newline
So let us assume that \( \alpha_0 = 0 \), the normalization (\ref{Conditions3}) holds and the initial datum is given by (\ref{Initial}): equation (\ref{Laplace}) then becomes
	\beq
	\label{Eq2}
		\tilde{q}(p) = - 4 \pi \sqrt{\frac{i}{p}} \: \underset{k \neq 0}{\sum_{k \in \mathbb{Z}}} \: \alpha_k \: \tilde{q}(p+i \omega k) - 2i \sqrt{\frac{2 \pi i}{p}} \: \frac{1 - \sqrt{-ip}}{1 + ip}
	\eeq
with the choice (\ref{Branch}) for the branch cut of \( \sqrt{p} \). By Proposition \ref{Analy} the solution is analytic on the open right half plane. In the following section we shall study the singularities on the imaginary axis.

\subsection{Singularities on the imaginary axis}

Setting \( q_n(p) \equiv \tilde{q}(p+i \omega n) \), \( p \in \mathcal{I} = [0,\omega) \), as in Section 3.1, equation (\ref{Eq2}) assumes the form (\ref{Eqr1}), 
	\beq
	\label{Eqr3}
		q(p) = \mathcal{M}(p) \:  q(p) + o(p)
	\eeq
		with
	\beq
		\big( \mathcal{M} q \big)_n (p) \equiv - \frac{4 \pi}{\sqrt{\omega n - ip}} \: \underset{k \neq 0}{\sum_{k \in \mathbb{Z}}} \: \alpha_k \: q_{n+k}(p)
	\eeq
and \( o(p) = \{ o_n(p) \}_{n \in \mathbb{Z}} \), 
	\beq
		o_n(p) \equiv - \frac{2i \sqrt{2 \pi}}{\sqrt{\omega n -ip}\:\:(1 + \sqrt{\omega n - ip})}
	\eeq

	\begin{pro}
	\label{Compact1}
		For \( p \in \mathcal{I} \), \( \Re(p) = 0 \), \( p \neq 0 \), \( \mathcal{M}(p) \) is an analytic operator-valued function and \( \mathcal{M}(p) \) is a compact operator on \( \ell_2(\mathbb{Z}) \).
	\end{pro}

	\emph{Proof:} 
		See the proof of Proposition \ref{Compact}.
		\begin{flushright} 
			\( \Box \)
		\end{flushright}
	
	\begin{pro}
	\label{Analyticity1}
		There exists a unique solution \( q_n(p) \in \ell_2(\mathbb{Z}) \) of (\ref{Eqr3}) and it is analytic on the imaginary axis for \( p \neq 0 \).
	\end{pro}
	
	\emph{Proof:} 
		See the proof of Proposition \ref{Analyticity}.
		\begin{flushright} 
			\( \Box \)
		\end{flushright}

	\begin{pro}
	\label{BranchPoints1}	
		If \( \{ \alpha_n \} \) satisfies (\ref{Conditions2}) and the genericity condition (\ref{Genericity}), the solution of equation (\ref{Eqr3}) has the form \( q_n(p) = c_n(p) + d_n(p) \sqrt{p} \), \( n \in \mathbb{Z} \), in a neighborhood of \( p = 0 \), where the functions \( c_n(p) \) and \( d_n(p) \) are analytic at \( p = 0 \).
	\end{pro}

	\emph{Proof:}
		Let us proceed as in the proof of Proposition \ref{BranchPoints}: setting \( q_n = r_n + t_n q_0 \), \( n \in \mathbb{Z} \setminus \{0\} \), where \( \{ t_n \} \) is the solution of 
		\beq
		\label{Eqtn1}
			t_n =  - \frac{4 \pi}{\sqrt{\omega n - ip}} \underset{k \neq 0, -n}{\sum_{k \in \mathbb{Z}}} \alpha_k  t_{n+k} - \frac{4 \pi \alpha_{-n}}{\sqrt{\omega n - ip}}
		\eeq
		A slightly different version of Lemma \ref{r_n} guarantees that the solution \( \{ t_n \} \in \ell_2(\mathbb{Z} \setminus \{0\}) \) is unique and analytic at \( p = 0 \).
		\newline
		By means of this substitution we obtain 
		\beq
		\label{Eqrn1}
			r_n =  - \frac{4 \pi}{\sqrt{\omega n - ip}} \underset{k \neq 0,-n}{\sum_{k \in \mathbb{Z}}} \alpha_k  r_{n+k} - \frac{2i \sqrt{2 \pi}}{\sqrt{\omega n -ip}\:\:(1 + \sqrt{\omega n - ip})}
		\eeq
		and
		\bdm
			q_0 =  - \frac{4 \pi}{\sqrt{- ip}} \underset{k \neq 0}{\sum_{k \in \mathbb{Z}}} \alpha_k  \big( r_k + t_k q_0 \big) - \frac{2i \sqrt{2 \pi}}{\sqrt{-ip}\:\:(1 + \sqrt{- ip})}
		\edm
		or 
		\bdm
			\big( \sqrt{- ip} + F \big) q_0 = G - \frac{2 \sqrt{2 \pi}}{1 + \sqrt{- ip}}
		\edm
		where (like in the proof of Proposition \ref{BranchPoints})
		\bdm
			F \equiv 4 \pi \underset{k \neq 0}{\sum_{k \in \mathbb{Z}}} \alpha_k t_k 
		\edm
		and 
		\bdm
			G \equiv - 4 \pi \underset{k \neq 0}{\sum_{k \in \mathbb{Z}}} \alpha_k r_k
		\edm
		Moreover \( F(0) \neq 0 \), because of genericity condition (\ref{Genericity}) (see the proof of Proposition \ref{BranchPoints}), \( F \) and \( G \) are analytic in a neighborhood of \( p = 0 \) (see Lemma \ref{r_n}), so that
		\bdm
			q_0 = F^{\prime} + \sqrt{p} \:\: G^{\prime}
		\edm
		where \( F^{\prime} \) and \( G^{\prime} \) are analytic and 
		\bdm
			G^{\prime} \equiv \frac{2 \sqrt{-2 \pi i} ( F + 1) - \sqrt{-i} (1+ip)G}{(1+ip)(F^2 + ip)} 
		\edm
		\begin{flushright} 
			\( \Box \)
		\end{flushright}

\subsection{Complete ionization in the generic case}

As in section 3 we can now state the main result:

	\begin{teo}
		If \( \{ \alpha_n \} \) satisfies (\ref{Conditions2}) and the genericity condition (\ref{Genericity}) with respect to \( \mathcal{T} \), as \( t \rightarrow \infty \),
		\beq
			|q(t)| \leq A \:  t^{-\frac{3}{2}} + R(t)
		\eeq
		where \( A \in \mathbb{R} \) and \( R(t) \) has an exponential decay, \( R(t) \sim C e^{-Bt} \) for some \( B > 0 \).
	\end{teo}
	
	\emph{Proof:} 
		See the proof of Theorem \ref{Decay}.
		\begin{flushright} 
			\( \Box \)
		\end{flushright}

	\begin{cor}
		If \( \{ \alpha_n \} \) satisfies (\ref{Conditions2}) and the genericity condition (\ref{Genericity}) with respect to \( \mathcal{T} \), the system shows asymptotic complete ionization and, as \( t \rightarrow \infty \),
		\bdm
			|\theta(t)| \leq D \: t^{-\frac{3}{2}} + E(t) 	
		\edm
		where \( D \in \mathbb{R} \) and \( E(t) \) has an exponential decay.
	\end{cor}
	
	\emph{Proof:} 
		See the proof of Corollary \ref{Survive}.
		\begin{flushright} 
			\( \Box \)
		\end{flushright}

	\begin{teo}
		If \( \{ \alpha_n \} \) satisfies (\ref{Conditions2}) and the genericity condition (\ref{Genericity}) with respect to \( \mathcal{T} \), every \( \Psi \in L^2(\mathbb{R}^3) \) is a scattering state of \( H_{\alpha(t)} \), i.e.
		\bdm
			\lim_{t \rightarrow \infty} \frac{1}{t} \int_0^t d\tau \: \big\| F(|\vec{x}| \leq R) U(\tau,0) \Psi \big\|^2 = 0
		\edm
		Moreover the discrete spectrum of the Floquet operator is empty.
	\end{teo}

	\emph{Proof:} 
		See the proof of Proposition \ref{DecayG} and Theorem \ref{Scattering}.
		\begin{flushright} 
			\( \Box \)
		\end{flushright}

\section{CASE III: \( \alpha_0 > 0 \)}

To complete the analysis of the problem, we shall consider the case of mean greater than \( 0 \): taking the normalization (\ref{Conditions3}) and the initial condition (\ref{Initial}), (\ref{Laplace}) assumes the form (\ref{Eq1}):
	\beq
	\label{Eq3}
		\tilde{q}(p) =  - \frac{4 \pi}{4 \pi \alpha_0 + \sqrt{-ip}} \: \underset{k \neq 0}{\sum_{k \in \mathbb{Z}}} \: \alpha_k \: \tilde{q}(p+i \omega k) - \frac{2i \sqrt{2 \pi}}{4 \pi \alpha_0 + \sqrt{-ip}} \frac{1 - \sqrt{-ip}}{1 + ip}
	\eeq
Analyticity of the solution on the open right half plane is a consequence of Proposition \ref{Analy}.
\newline
Moreover, following the discussion contained in section 3 and setting \( q_n(p) \equiv \tilde{q}(p+i \omega n) \), \( \Im(p) \in [0,\omega) \), the equation assumes the form (\ref{Eqr1}).
\newline
Let us now consider the behavior on the imaginary axis: singularities for \( \Re(p) = 0 \) are associated to zeros of \( 4 \pi \alpha_0 + \sqrt{\omega n + s} \), \( s \in [0, \omega) \), but, since \( \alpha_0 > 0 \), it is clear that the expression can not have zeros on the imaginary axis. Hence the proof of Proposition \ref{Analyticity} can be extended to the closed right half plane except the origin:
	
	\begin{pro}
	\label{Analyticity2}
		If \( \{ \alpha_n \} \) satisfies (\ref{Conditions2}), the solution \( \tilde{q}(p) \) of (\ref{Eq3}) is unique and analytic for \( \Re(p) \geq 0 \), \( p \neq i \omega n \), \( n \in \mathbb{Z} \).
	\end{pro}
	
	\emph{Proof:}
		See the proof of Proposition \ref{Analyticity}, Propositions \ref{Compact} and \ref{Homogeneous} and the previous discussion. 
		\begin{flushright} 
			\( \Box \)
		\end{flushright}

Moreover the behavior at the origin is described by the following
	
	\begin{pro} 
	\label{BranchPoints2}
		If \( \{ \alpha_n \} \) satisfies (\ref{Conditions2}) and the genericity condition with respect to \( \mathcal{T} \) (\ref{Genericity}), then, in an imaginary neighborhood of \( p = i \omega n \), \( n \in \mathbb{Z} \), the solution of equation (\ref{Eq3}) has the form \( \tilde{q}(p) = c_n(p) + d_n(p) \sqrt{p- i \omega n} \), where the functions \( c_n(p) \) and \( d_n(p) \) are analytic at \( p = i \omega n \).
	\end{pro}

	\emph{Proof:}
		The proof of Proposition \ref{BranchPoints} still applies with only one difference: since, independently on \( \omega \), the solution can not have a pole on the imaginary axis, we need not to distinguish between the resonant case and the non-resonant one.
		\begin{flushright} 
			\( \Box \)
		\end{flushright}

We can now prove asymptotic complete ionization of the system:

	\begin{teo}
		If \( \{ \alpha_n \} \) satisfies (\ref{Conditions2}) and the genericity condition (\ref{Genericity}) with respect to \( \mathcal{T} \), as \( t \rightarrow \infty \),
		\beq
			|q(t)| \leq A \:  t^{-\frac{3}{2}} + R(t)
		\eeq
		where \( A \in \mathbb{R} \) and \( R(t) \) has an exponential decay, \( R(t) \sim C e^{-Bt} \) for some \( B > 0 \).
		\newline
		Moreover the system shows asymptotic complete ionization and, as \( t \rightarrow \infty \),
		\bdm
			|\theta(t)| \leq D \: t^{-\frac{3}{2}} + E(t) 	
		\edm
		where \( D \in \mathbb{R} \) and \( E(t) \) has an exponential decay.
	\end{teo}
	
	\emph{Proof:} 
		See the proof of Theorem \ref{Decay} and Corollary \ref{Survive}.
		\begin{flushright} 
			\( \Box \)
		\end{flushright}

	\begin{teo}
		If \( \{ \alpha_n \} \) satisfies (\ref{Conditions2}) and the genericity condition (\ref{Genericity}) with respect to \( \mathcal{T} \), every \( \Psi \in L^2(\mathbb{R}^3) \) is a scattering state of \( H_{\alpha(t)} \), i.e.
		\bdm
			\lim_{t \rightarrow \infty} \frac{1}{t} \int_0^t d\tau \: \big\| F(|\vec{x}| \leq R) U(\tau,0) \Psi \big\|^2 = 0
		\edm
		Moreover the discrete spectrum of the Floquet operator is empty.
	\end{teo}

	\emph{Proof:} 
		See the proof of Proposition \ref{DecayG} and Theorem \ref{Scattering}.
		\begin{flushright} 
			\( \Box \)
		\end{flushright}
\textbf{Remark:} If \( \alpha(t) \geq 0 \), \( \forall t \in \mathbb{R}^+ \), Proposition \ref{BranchPoints2} holds without the genericity condition on the Fourier coefficients of \( \alpha(t) \): for instance the genericity condition enters (see the proof of Proposition \ref{BranchPoints}) in the proof of absence of non-zero solutions of the homogeneous equation 
	\bdm
		t_n = - \frac{4 \pi}{4 \pi \alpha_0 + \sqrt{\omega n + s}} \: \underset{k \neq 0,-n}{\sum_{k \in \mathbb{Z}}} \: \alpha_k \: t_{n+k}
	\edm
where \( s \in [0,\omega) \). Let us suppose that there exists a non-zero solution \( \{ T_n \} \in \ell_2(\mathbb{Z}) \). Multiplying both sides of the equation by \( T_n^* \), one has
	\bdm
		\underset{n \neq 0}{\sum_{n \in \mathbb{Z}}} \sqrt{\omega n + s} \: | T_n |^2 = - 4 \pi \underset{n,k \neq 0}{\sum_{n,k \in \mathbb{Z}}} T_n^* \alpha_{k-n}  T_{k}
	\edm
Since the right hand side is real, \( T_n = 0 \), \( \forall n < 0 \). Moreover, fixing \( T_0 = 0 \) and setting
	\bdm
		T(t) \equiv \sum_{n \in \mathbb{Z}} T_n \: e^{-i \omega n t}
	\edm
it follows that
	\bdm
		- 4 \pi \sum_{n,k \in \mathbb{Z}} T_n^* \alpha_{k-n}  T_{k} = - 4 \pi \Big( T(t), \alpha(t) T(t) \Big)_{L^2([0,T])} \leq 0
	\edm
because \( \alpha(t) \geq 0 \), \( \forall t \in [0,T] \), but the left hand side is positive and then \( Q_n = 0 \), \( \forall n \in \mathbb{Z} \).

\section{Conclusions and Perspectives}

In sections 3,4 and 5 we have proved that, under the genericity condition on \( \alpha(t) \), the system defined in section 2 shows asymptotic complete ionization, independently on its frequency. 
\newline
If \( \inf(\alpha(t)) < 0 \), the genericity condition may be a necessary condition to have complete ionization: for example, in one dimension, it is possible to exhibit (see \cite{Cost1}) explicit functions \( \alpha(t) \) for which the genericity condition fails\footnote{A simple example of \( \alpha(t) \), for which the genericity condition is not satisfied is the geometric series, \( \alpha_n = \lambda^{|n|} \) for some \( \lambda < 1 \).} and the ionization is not complete. On the other hand, also in one dimension, it is not known whether the condition is necessary. It would be interesting to check if non generic \( \alpha(t) \) gives rise to asymptotic partial ionization in three dimensions.
\newline
A possible way to investigate this problem is the analysis of the discrete spectrum of the Floquet operator. If one can find an explicit relation between existence of eigenvalues of the Floquet operator and the genericity condition, it would be probably easy to check if the condition is truly necessary.
\newline
On the other hand, as we expected, if \( \alpha(t) \) is positive at any time, no further condition on \( \alpha(t) \) is required to prove complete ionization. 
\newline
Two interesting future applications of these methods can be the problem of complete ionization for moving point interactions and for \( N \) time-dependent point interactions. Indeed there are simple examples in which asymptotic complete ionization occurs also for moving sources (see \cite{Corr1}). 
\newline
\mbox{}	\\
\mbox{}	\\
\textbf{Acknowledgments:} M.C. is very grateful to Prof. Ludwik Dabrowski  and the INTAS Research Project nr. 00-257 of European Community, ``Spectral Problems for Schr\"{o}dinger-Type Operators'', for the support.

\end{document}